\documentstyle[aps,pre,eqsecnum,12pt]{revtex}

\begin{document}
\tighten
\title{Conservation of circulation in magnetohydrodynamics }

 \author{ Jacob D. Bekenstein\thanks{Electronic mail:
     bekenste@vms.huji.ac.il} and Asaf
Oron\thanks{Electronic mail:asafo@alf.fiz.huji.ac.il}}

\address{\it The Racah Institute of Physics, Hebrew University of
Jerusalem,\\ Givat Ram, Jerusalem 91904, Israel}

\date{\today}

\maketitle

\begin{abstract}
We demonstrate, both at the Newtonian and (general) relativistic levels, the
existence of a generalization of Kelvin's circulation theorem (for pure fluids)
which is applicable to perfect magnetohydrodynamics.  The argument is based on
the least action principle for magnetohydrodynamic flow.  Examples of the new
conservation law are furnished. The new theorem should be helpful in
identifying new kinds of vortex phenomena distinct from magnetic ropes or
fluid vortices.
\end{abstract}
 \section{Introduction}

Kelvin's theorem on the conservation of circulation of a simple perfect
fluid has played an important role in the development of hydrodynamics.  For
instance, it shows that potential flows are possible, that isolated
vortices can exist, that they obey the Helmholtz laws, etc.  Kelvin's
theorem is valid only for flows in which the body force per unit mass is a
gradient; mostly this includes incompressible or isentropic flows of
one--component fluids.

Most flows in geophysics and astrophysics are more complicated.  In
particular, many fluids in the real world carry magnetic fields: they are
magnetofluids.   Yet the Lorentz force per unit mass on a
magnetofluid is almost {\it never\/} a perfect gradient.  Thus the circulation
theorem in its original form is almost never true in magnetohydrodynamics
(MHD).  Must we then surrender the many insights that Kelvin's theorem
conferred on pure hydrodynamics ? 

Not necessarily. One might speculate that a suitable combination of fluid
velocity ${\bf v}$ and magnetic induction ${\bf B}$ may inherit the
property of having a ``circulation'' on a closed curve which is preserved as
that curve is dragged with the magnetofluid.  Such conserved circulation
might play as useful a role in MHD as has Kelvin's circulation in pure
fluid dynamics.  For example, it might help characterize a set of
magnetoflows as being potential in some sense, with consequent
simplification of this intricate subject.  Or it might help to characterize
a new type of vortex, a hybrid vorticity--magnetic rope.  In view of the
importance of the vortex phenomenon in contemporary physics, this last
possibility is by itself ample reason to delve into the subject.

Two decades ago, E. Oron\cite{bekenstein1} discovered, with the formalism
of relativistic perfect MHD, a circulation theorem of the above kind.
Although some of its consequences for new helicity conservation laws have
been explored\cite{bekenstein2}, this new conserved circulation has remained
obscure.  Contributing to this, no doubt, is the fact that it has only
been derived relativistically, and that this derivation is an intricate one,
even for relativistic MHD.  In addition, Oron's derivation assumes both
stationary symmetry and axisymmetry, while it is well known that Kelvin's
theorem requires neither of these.

In the present paper we use the least action principle to give a rather
straightforward  existence proof for a generically conserved hybrid
velocity--magnetic field circulation within the framework of perfect MHD
which does not depend on spacetime symmetries.  We do this both at the
Newtonian (Sec.II) and general relativistic (Sec.III) levels; the importance
of MHD effects in pulsars, active galactic nuclei and cosmology underscores
that this last arena is not just of academic importance.

As mentioned, we approach the whole problem not from equations of motion,
but from the least action principle.  Lagrangians for nonrelativistic pure
perfect flow have been proposed by Herivel\cite{herivel},
Eckart\cite{eckart}, Lin\cite{lin}, Seliger  and  Witham\cite{seliger},
Mittag, Stephen and Yourgrau\cite{mittag} and others.  Many of the proposed
Lagrangians necessarily imply irrotational flow, {\it i.e.\/} not to
generic flow, a deficiency which is often missed by the authors.  
Lin\cite{lin} introduced a device that allows vortical flows to be
encompassed.  This device is used by Seliger  and  Witham.  Lagrangians
for nonrelativistic perfect MHD flow in Eulerian coordinates have been 
proposed by Eckart\cite{eckart2}, Henyey\cite{henyey}, Newcomb\cite{newcomb},
Lundgren\cite{lundgren} and others.

In special relativity Penfield\cite{penfield} proposed a perfect
fluid Lagrangian which admits vortical isentropic flow.  The early
general relativistic Lagrangian of Taub\cite{taub1,taub2} as well
as the more recent one by Kodama et. al\cite{kodama} describe only
irrotational perfect fluid flows. The Lin device is incorporated by
Schutz\cite{schutz}, whose perfect fluid Lagrangian admits vortical as
well as irrotational  flows in general relativity. Carter
\cite{carter1,carter2} introduced Lagrangians for particle-like motions from
which can be inferred the properties of fluid flows, including vortical
ones.  Achterberg\cite{achterberg} proposed a general relativistic MHD action,
which, however, describes only ``irrotational'' flows. Thompson
\cite{thompson} used this Lagrangian in the extreme relativistic
limit. Heyl and Hernquist \cite{heyl}  modified it to include QED
effects. In this paper we follow mostly Seliger  and  Witham\cite{seliger} 
and Schutz\cite{schutz}.

In Sec.~II.A we propose a nonrelativistic MHD Lagrangian, and show in
Sec.~II.B and II.C that it gives rise to the correct equations of motion for
the density, entropy, velocity and magnetic fields in Newtonian MHD. 
In Sec.~II.D we derive from it the conserved circulation, defined in terms
of a new vector field ${\bf R}$, and discuss its invariance under
redefinition of ${\bf R}$.  Sec.~II.E furnishes two examples of the
conserved circulation in action.  In Sec.~III.A we collect all the equations
of motion of general relativistic MHD, and propose a general relativistic
MHD Lagrangian in Sec.~III.B. Secs.~III.C and III.D recover all the
relativistic MHD equations of motion from it.  Finally in Sec.~III.E we
generalize the conserved MHD circulation to the general relativistic case.  

\section{Variational Principle in Eulerian Coordinates}

\subsection{The Lagrangian Density}

Perfect MHD describes situations where the flow is nondissipative, and, in
particular, when the magnetoflow  has ``infinite conductivity'', and where
Maxwell's displacement current may be neglected in Ampere's equation.  We
shall adopt this approximation.  We work in eulerian coordinates: all
physical quantities are functions of coordinates $x_i$ or $ {\bf r}$ which
describe a fixed point in space.  We first summarize the MHD equations.  We
work in units for which $c=1$.

First of all, the fluid obeys the equation of continuity ($\partial_t\equiv
\partial/\partial t)$
\begin{equation}
\label{continu}
\partial_t \rho +\nabla \cdot \left( \rho {\bf v}\right) =0,
\end{equation}
where $\rho( {\bf r}, t)$ is the mass density per unit volume of the fluid
and ${\bf v}( {\bf r}, t)$ is the fluid's velocity field.  Second, since
there is no dissipation, $s$, the entropy per unit mass, must be conserved
along the flow:
\begin{equation}
\label{adiabatic0}
Ds\equiv\partial_t s +{\bf v}\cdot\nabla s =0.
\end{equation}
Here we have defined the convective derivative $D$, which in Cartesian
coordinates has the same form for scalars or vectors. With the help of
Eq.~(\ref{continu}) this equation can be written as
\begin{equation}
\label{adaibatic1}
\partial_t {(\rho s)}+\nabla \cdot \left( \rho s{\bf v}\right) =0.
\end{equation}
Third, ``infinite conductivity" implies that ${\bf E} + ({\bf v}/c) \times
{\bf B}=0$, where ${\bf E}$ and ${\bf B}$ are the electric and magnetic
fields, respectively.  Combining this with Faraday's equation yields the so
called field-freezing equation
\begin{equation}
\label{ff}
\partial_t {\bf B}=\nabla \times \left( {\bf v}\times {\bf B}\right),
\end{equation}
which implies Alfven's law of conservation of the magnetic flux
through a closed loop moving with the flow.
Finally, the evolution of the velocity field is governed by the MHD Euler
equation,
\begin{equation}
\label{mag_euler}
\rho D{\bf v}=-\nabla p-\rho\nabla U +{(\nabla\times{\bf B})\times{\bf
B}\over 4\pi},
\end{equation}
where $p$ is the fluid's pressure (here assumed isotropic), and $U({\bf r},t)$
is the gravitational potential.

The least action principle is in general
\begin{equation}
    \delta S[f_a] \equiv \delta \int dt \int d^3r\, {\cal L}(f_a,\partial_t
f_a,\nabla f_a) =0.
\label{action}
\end{equation}
Here the action $S$ is a functional of various fields $f_a({\bf r},t)$,
$a=1, 2, \cdots$. One varies each $f_a$, transfers time and space
derivatives of each variation $\delta f_a$ to the adjacent factor by
integration by parts, and sets to zero the overall coefficient of the bare
$\delta f_a$.  This gives us the Lagrange--Euler equation
\begin{equation}
\partial_t\left({\partial {\cal L}\over \partial (\partial_t
f_a)}\right) + \nabla\cdot \left({\partial {\cal L}\over \partial \nabla
f_a}\right) -{\partial {\cal L}\over \partial f_a}=0.
\label{EL}
\end{equation}
It is usually more convenient to get the equation for each $f_a$ {\it ab
initio\/} by the above procedure, rather than by using Eq.~(\ref{EL}).

 We now propose the following Lagrangian {\it density\/} for
MHD flow of perfect infinitely conducting  fluid which incorporates
Eqs.(\ref{continu}-\ref{ff}), as three Lagrange constraints
\begin{eqnarray}
{\cal L}&=&\rho {\bf v}^{2}/2-\rho
\epsilon
\left(
\rho ,s\right) -\rho U -{\bf B}^{2}/(8\pi) +
\nonumber \\
&+&\phi \left[ \partial_t\rho +\nabla \cdot \left[ \rho
{\bf v}\right) \right]
+  \eta \left[ \partial_t{(\rho s)}+\nabla \cdot
\left(\rho s{\bf v}\right)
                   \right]
\nonumber\\
&+&\lambda \left[ \partial_t{(\rho\gamma)} +\nabla \cdot
\left(
\rho
\gamma {\bf v}\right)
                   \right]
+{\bf K}\cdot \left[\partial_t{\bf B} -\nabla \times
\left( {\bf v}\times {\bf B}\right) \right].
\label{le1}
\end{eqnarray}
In the above $\epsilon \left(\rho,s\right)$ is the thermodynamic
internal energy per unit mass; in the total Lagrangian the corresponding
total internal energy enters as a potential energy.  The magnetic energy, the
volume integral of ${\bf B}^2/(8\pi)$, also enters the total Lagrangian as a
potential energy.

In Eq.~(\ref{le1})  $\phi$, $ \eta $ are Lagrange
multiplier fields which locally enforce the conservation laws
(\ref{continu}-\ref{adiabatic0}), as may be verified by varying with respect
to these multipliers.   ${\bf K}$ is a triplet of Lagrange multiplier fields
which enforce the field--freezing constraint Eq.~(\ref{ff}): varying
with respect to ${\bf K}$ reproduces Eq.~(\ref{ff}) at every point and time.
Finally, $\lambda$ is a Lagrange multiplier field which enforces the Lin
constraint on a new field, $\gamma$:
\begin{equation}
\partial_t(\rho\gamma)+\nabla\cdot(\rho{\bf v}\gamma) = 0\qquad {\rm
or}\qquad D\gamma = 0.
\label{Lin}
\end{equation}
Here we have used Eq.~(\ref{continu}) to reduce to the second form.
Lin's field $\gamma$, like $s$, is conserved along the flow, but
unlike $s$ it does not occur elsewhere in the Lagrangian.  Lin interprets
$\gamma({\bf r}, t)$ as one of the three initial {\it Lagrangian\/} coordinates
which label each fluid element.  But whatever the interpretation, the
condition~(\ref{Lin}) is essential so that the flow can be vortical also
in the limit ${\bf B}\rightarrow 0$. This matter is further discussed in
the following section.

\subsection{The Equations of Motion}

Can our proposed Lagrangian density reproduce all the equations of motion of
perfect MHD flow ?  We have already seen that it does reproduce
Eqs.~(\ref{continu}-\ref{adiabatic0}) and (\ref{ff}).  Let us now vary
$\gamma$ to get
\begin{equation}
\label{le4}
D\lambda = 0,
\end{equation}
so that $\lambda$, like $\gamma$, is conserved with the flow.  Both this and
Eq.~(\ref{Lin}) will be essential in demonstrating the existence of the new
conserved circulation.  Next we vary
$s$; remembering that $(\partial \epsilon/ \partial
s)_\rho$ is just the fluid's temperature $T$, we have
\begin{equation}
\label{le6}
D\eta=-T,
\end{equation}
which establishes that $\eta$ decreases along the flow.  The next variation
is one with respect to $\rho$. Recalling that $(\partial \epsilon/ \partial
\rho)_s=p/\rho^2$, introducing the enthalpy per unit mass
$w=\epsilon+p/\rho$, and using Eqs.~(\ref{le4}-\ref{le6}) we get
\begin{equation}
\label{le7}
D\phi=v^{2}/2-w+T- U.
\label{phi}
\end{equation}

When we vary ${\bf v}$ in the action we may take advantage of the identity
$\nabla\cdot({\bf A}\times{\bf B})={\bf B}\cdot\nabla\times {\bf A}-{\bf
A}\cdot\nabla\times {\bf B}$ and Gauss' theorem to flip the curl operation
from $\delta{\bf v}\times {\bf B}$ onto ${\bf K}$.  Then the identity ${\bf
A}\cdot{\bf B}\times{\bf C}=-{\bf B}\cdot{\bf A}\times{\bf C}$ helps to
shift the $\delta{\bf v}$ into the position of a factor in a scalar product.
We may then factor out the common $\delta{\bf v}$ and isolate the vector
equation
\begin{eqnarray}
\label{le3}
{\bf v}&=&\nabla \phi +\gamma \nabla \lambda +s\nabla \eta + {\bf Q}
\\
\label{me3}
{\bf Q} &\equiv& {\bf B}\times{\bf R}/\rho.
\end{eqnarray}
where $ {\bf R}\equiv \nabla \times {\bf K}$. This is neither a solution for
${\bf v}$ ($\lambda$ and $\eta$ not known), nor an equation of motion (${\bf
v}$ appears undifferentiated).  In the next subsection we show that this
prescription for ${\bf v}$ leads to the MHD Euler equation (\ref{mag_euler}).

Expression (\ref{le3}) shows the importance of including Lin's field
$\gamma$. For suppose we consider an unmagnetized fluid in isentropic ($s=$
const.) flow.  Without $\gamma$ the expression for ${\bf v}$ is a perfect
gradient, which means the proposed Lagrangian density describes only
irrotational flows, a small subset of all possible ones. It is well known
\cite{seliger,mittag} that this problem does not appear when one
couches the problem in Lagrangian coordinates because one gets then an
equation, not for  $ {\bf v}$, but for the fluid's acceleration.
Lin's\cite{lin} way out of this difficulty is to remember that
the initial coordinates of the fluid element are maintained throughout its
flow. These coordinates ``label'' the element, and this can be interpreted
as a triplet of constraints (one for each coordinate) of the form $
\lambda _i\left( {\partial b_i}/{\partial t}+\nabla b_i\right),  $ where
${\bf b}$ is the initial vector coordinate for the element in question.
Lundgren\cite{lundgren} used this triplet form for the MHD case. It was
later shown (see for example\cite{seliger}) that the triplet can be reduced
to a single constraint with the help of Pfaff's theorem.  One thus returns
to form (\ref{le1}) of the Lagrangian density and Eq.~(\ref{le3}) for the
fluid velocity. The vorticity is now (still excluding ${\bf B}$)
\begin{equation}
\label{le13}
{\bf \omega }=\nabla \times {\bf v}= \nabla \gamma
\times \nabla \lambda +\nabla s\times \nabla \eta,
\end{equation}
so we see that isentropic vortical flow is possible.

In the MHD case, the magnetic term in Eq.~(\ref{le3}) contributes to the
vorticity.  Henyey\cite{henyey}, who suggested a Lagrangian density similar
to ours, occasionally dropped the Lin term in the MHD case.   However, we shall
retain the Lin term throughout.  It might seem peculiar at first that adding a
constraint like Lin's permits the appearance of solutions (vortical) which
were forbidden before it was imposed. But we must remember that we add  to the
Lagrangian not only a constraint, but also a new degree of freedom,
$\gamma({\bf r},t)$, and it is natural that with more degrees of freedom the
class of allowed flows will expand.

Finally, we vary ${\bf B}$ in the action; by similar manipulation to those
which gave Eq.~(\ref{le3}) we get
\begin{equation}
\label{le8}
\partial_t{\bf K}={\bf v} \times {\bf R}
-{\bf B}/(4\pi).
\end{equation}
Taking the curl of this equation we get the more convenient one
\begin{equation}
\partial_t{\bf R}=\nabla \times [ {\bf v}
\times {\bf R}-{\bf B}/(4\pi)]=\nabla \times ( {\bf v}\times {\bf R})-{\bf J}.
\label{dR}
\end{equation}
Here ${\bf J}= \nabla \times {\bf B} / 4 \pi $ is the
electric current density coming from Ampere's equation. Notice the similarity
between Eq.~(\ref{dR}) and~(\ref{ff}). Eq.~(\ref{dR}) says that
the rate of change of the flux of ${\bf R}$  through the surface spanning a
closed curve carried with the flow equals minus the flux of the
electric current density through that curve.

\subsection{The MHD Euler equation}

We now show that the Lagrangian density (\ref{le1}) yields the correct
MHD Euler equation. We first operate with the convective derivative $D$ on
Eq.~(\ref{le3}) remembering that $Ds=0$ and $D\gamma=0$:
\begin{equation}
\label{me0}
D{\bf v}=D\nabla\phi+\gamma D\nabla\lambda+sD\nabla\eta +D{\bf Q}.
\end{equation}
We now use the identity
\begin{equation}
\label{me4}
D\nabla=\nabla D - (\nabla{\bf v}) \cdot \nabla,
\end{equation}
where in Cartesian coordinates
\begin{equation}
\label{tensor}
[(\nabla{\bf v}) \cdot \nabla]_i \equiv \sum_j {\partial v_j\over \partial
x_i}{\partial\over\partial x_j},
\label{grad_v}
\end{equation}
in conjunction with Eqs.~(\ref{le4}-\ref{le7}) to transform Eq.~(\ref{me0})
into
\begin{eqnarray}
D{\bf v}&=&\nabla (v^{2}/2-w+Ts-U)-s\nabla  T
-s(\nabla{\bf v})\cdot \nabla\eta
\nonumber
\\
\label{me5}
&-&(\nabla{\bf v}) \cdot \nabla\phi-\gamma(\nabla{\bf v}) \cdot
\nabla\lambda + D{\bf Q}.
\end{eqnarray}

From the thermodynamic identity $dw = T ds + dp/\rho$ we infer
\begin{equation}
\label{me7}
-\nabla w + T\nabla s = -\nabla p/\rho,
\end{equation}
and we also have $\nabla v^2/2 =
(\nabla{\bf v})\cdot {\bf v}$, where the meaning of the right hand side is
clear by analogy with Eq.~(\ref{grad_v}).  Thus Eq.~(\ref{me5}) turns into
\begin{equation}
D{\bf v} = -\nabla p/\rho -\nabla U + (\nabla{\bf v})\cdot({\bf v} -\nabla
\phi- s\nabla \eta -\gamma\nabla\lambda) + D{\bf Q}.
\label{semifinal}
\end{equation}
Finally comparing with Eq.~(\ref{le3})  we see that the last brackets stand
for ${\bf Q}$ so that
\begin{equation}
D{\bf v} = -\nabla p/\rho -\nabla U + (\nabla{\bf v})\cdot{\bf Q} + D{\bf
Q}.
\label{final}
\end{equation}
Thus, magnetic term aside, we have recovered the Euler equation
(\ref{mag_euler}).

We now go on to calculate the ${\bf Q}$ dependent terms. We may rewrite
the equation of continuity  (\ref{continu}) as
\begin{equation}
\label{me9}
D\rho=-\rho \nabla \cdot {\bf v},
\end{equation}
With this, the Gauss law $\nabla\cdot{\bf B}=0$ and the identity
$\nabla\times({\bf A}\times{\bf B})={\bf B}\cdot\nabla{\bf A}-{\bf
B}\nabla\cdot{\bf A}-{\bf A}\cdot\nabla{\bf B}+{\bf
A}\nabla\cdot{\bf B}$, Eq.~(\ref{ff}) may be recast in the well known form
\begin{equation}
\label{Btransport}
D({\bf B}/\rho)=\left(({\bf B}/\rho)\cdot\nabla\right) {\bf v}.
\end{equation}
Analogously, because $\nabla\cdot{\bf R}=0$, Eq.~(\ref{dR}) may be put in
the form
\begin{equation}
\label{Rtransport}
D{\bf R}=({\bf R}\cdot\nabla) {\bf v}-{\bf R}\nabla\cdot{\bf v}-{\bf J}.
\end{equation}
Therefore,
\begin{eqnarray}
\nonumber
(\nabla{\bf v})\cdot{\bf Q} + D{\bf
Q}&=& -(\nabla{\bf v})\cdot({\bf R}\times{\bf B}/\rho) -D{\bf R}\times{\bf
B}/\rho-{\bf R}\times D({\bf B}/\rho)
\\
&=& -(\nabla{\bf v})\cdot({\bf R}\times{\bf B}/\rho) -\left(({\bf
R}\cdot\nabla){\bf v}\right)\times({\bf B}/\rho)
\nonumber
\\
&+& (\nabla\cdot{\bf v}){\bf
R}\times({\bf B}/\rho) -{\bf R}\times\left(({\bf
B}/\rho)\cdot\nabla\right){\bf v}+{\bf J}\times{\bf B}/\rho.
\label{distrib}
\end{eqnarray}

The four terms in the second version of Eq.~(\ref{distrib}) involving
derivatives of ${\bf v}$ can be shown to cancel out by
expanding them out in Cartesian coordinates.  Hence, Eq.~(\ref{final}) is the
magnetic Euler equation with the usual Lorentz force per unit mass, ${\bf
J}\times{\bf B}/\rho$, in addition to the pure fluid terms. The fact that we
obtain the correct MHD equations (\ref{continu}-\ref{adiabatic0}), (\ref{ff})
and (\ref{mag_euler}) is testament to the correctness of our proposed
Lagrangian density Eq.~(\ref{le1}).  Note that Lin's field $\gamma$ has
disappeared from the final equation of motion.

\subsection{Circulation Conservation Law}
\label{circ}

With the help of  the above formalism, we can now prove the existence of
a generalization of Kelvin's circulation theorem applicable to perfect MHD. 
Let us calculate the line integral of the vector
\begin{equation}
\label{mke1}
{\bf Z}={\bf v}+{\bf R}\times{\bf B}/\rho
\end{equation}
along a closed curve $ {\cal C} $ drifting with the fluid:
\begin{equation}
\label{mke2}
\Gamma =\oint _{{\cal C}}{\bf Z}\cdot d  {\bf r}.
\end{equation}
According to Eq.~(\ref{le3}) this integral is
\begin{equation}
\label{mke3}
\Gamma =\oint_{{\cal C}}\nabla \phi\cdot d {\bf r} +\oint _{{\cal
C}}\gamma \nabla \lambda \cdot d  {\bf r}+
\oint _{{\cal C}}s\nabla \eta\cdot d {\bf r}.
\end{equation}
The term involving $\phi$ obviously vanishes (we assume all the Lagrange
multipliers are single valued). For like reason so does the term involving
$\eta$ in the isentropic ($s=$ const ) case as
$s$ can be taken out of the integral.  The
middle integral can be written  $\oint_{{\cal C}}\gamma\,d\lambda$,
where $d\lambda\equiv \nabla\lambda\cdot d{\bf r}$.  But
Eqs.~(\ref{Lin}-\ref{le4}) tell us that both
$\gamma$ and $\lambda$ are conserved along the flow.  Hence $\Gamma$ remains
constant as ${\cal C}$ drifts along with the flow.  Since in the limit
${\bf B}\rightarrow 0$, $\Gamma$ becomes Kelvin's circulation, we
have found an extension of Kelvin's theorem to perfect MHD.  Obviously
the conservation of $\Gamma$ implies the conservation of the flux of $\nabla
\times {\bf Z} $ through ${\cal C}$.

The vector field ${\bf R}$ is not unique for a given physical situation. For
example, the  change ${\bf R}\rightarrow {\bf R}+ k {\bf B}$ ($k$ a real
constant) leaves invariant all equations of motion,
Eqs.~(\ref{Lin}-\ref{me3}), (\ref{dR}), and (\ref{final}), as well as the
conserved circulation expressions (\ref{mke1}-\ref{mke2}).  In addition, 
suppose that at time
$t=0$ we define an arbitrary solenoidal (divergence--free) field
${\bf b}$ all over the flow, and then evolve it in time  as a passive
vector, i.e., in accordance with the frozen--in field equation (\ref{ff}). 
Comparing with Eq.~(\ref{dR}) we see that ${\bf R}+k{\bf b}$ and ${\bf R}$
obey the same equation, and both are permanently solenoidal [this property is
obviously preserved by Eqs.~(\ref{ff}) and (\ref{dR}) in the MHD
approximation].  

If in ${\bf Z}$ we use ${\bf R}+k {\bf
b}$ in lieu of ${\bf R}$ to construct the conserved circulation,  $\Gamma$
gets the additional contribution
\begin{equation}
\Delta\Gamma=k\oint _{\cal C}\left({\bf b}\times {\bf
B}/\rho\right)
\cdot d {\bf r} =k \oint _{{\cal C}} {\bf B}\cdot \left(d{\bf r}\times{\bf
b/\rho}\right)
\label{DGamma}
\end{equation}
Here we have used a well known vector identity.   Now by analogy with ${\bf
B}$, ${\bf b}$ obeys Eq.~(\ref{Btransport}) which tells us that any two
elements of the fluid permanently lie on one and the same line of ${\bf
b}/\rho$, and their distance, if small, is proportional to  $|{\bf
b}|/\rho$\cite{LL}.  We can always make ${\bf b}$ small.  Then $d{\bf
r}\times{\bf b/\rho}$ is a vectorial element of area of a narrow closed
strip carried along by the fluid, one of whose edges coincides with ${\cal
C}$.  The integral in Eq.~(\ref{DGamma}) is just the flux of magnetic
induction through this strip (not through the space bounded by the strip),
and we know this is conserved by virtue of Alfven's law.  

Thus with the change ${\bf R}\rightarrow {\bf R}+k{\bf b}$
we added some conserved magnetic flux to $\Gamma$, and did not get a new
conserved circulation.  The MHD flow $\{ {\bf B}, {\bf v}, \rho, p\}$ is
evidently unchanged because the MHD Euler equation (\ref{mag_euler}) does
not contain ${\bf R}$, so we must conclude that in the expression for ${\bf
v}$, Eq.~(\ref{le3}-\ref{me3}), the change of the ${\bf Q}$ term
must be compensated by suitable changes in the Lagrange multipliers
$\phi+s\eta$ and $\lambda$ (recall that we are working with $s={\rm
const.}$).  Indeed, the initial choice of ${\bf b}$ involves a choice of two
functions because of the $\nabla\cdot{\bf b}=0$ constraint, so that the two
functions $\phi+s\eta$ and $\lambda$ are just enough to absorb the change
${\bf R}\rightarrow {\bf R}+k {\bf b}$ thus
generated  and leave ${\bf v}$ unchanged.   It is not possible to eliminate
${\bf R}$ altogether by the change ${\bf R}\rightarrow {\bf R}+k {\bf b}$
because ${\bf R}$ and ${\bf b}$ obey different equations.  This means the
circulation conservation law we have found cannot be reduced to an Alfven
type law; it is a new law.  

In Sec.\ref{last_sec} we shall discuss the freedom inherent in
${\bf R}$ by a covariant procedure.  Fixing the freedom is a necessary
step in any attempt to exhibit explicitly the conserved circulation.

\subsection{Examples}

First consider a situation where the fluid is isentropic but not flowing:
${\bf v}=0$. It follows from Eq.~(\ref{continu}) that $\rho=\rho_0({\bf r})$, 
and from Eq.~(\ref{ff}) that ${\bf B} = {\bf B}_0({\bf r})$. From these facts
and Eq.~(\ref{dR}) we see that
\begin{equation}
{\bf R}=-t\ \nabla\times{\bf B}_0({\bf r})/(4\pi)+ {\bf R}_0({\bf r}).
\end{equation}
Although the physical quantities are stationary, ${\bf R}$ is not.  This is
so because like the electromagnetic potential, ${\bf R}$ is not a measurable
quantity, being subject to ``gauge changes'' ${\bf R}\rightarrow {\bf R}+{\bf
b}$ as already discussed. According to Eq.~(\ref{mke1}) the conserved
circulation (around a contour fixed in space because ${\bf v}=0$) should be
\begin{equation}
\Gamma= -t\oint _{{\cal C}} {(\nabla\times{\bf B}_0)\times{\bf B}_0\over
4\pi\rho_0}\cdot d{\bf r} +
\oint _{{\cal C}} {{\bf R}_0\times {\bf B}_0\over 4\pi\rho_0}\cdot d{\bf r}.
\end{equation}
On the face of it, the time dependence of the first term in this
simple situation puts the claimed circulation conservation law in jeopardy.
However, according to the magnetic Euler equation (\ref{mag_euler}), the
first integrand here is equal to $\nabla U+\nabla p/\rho_0$ which, by
virtue of Eq.~(\ref{me4}) and the isentropic nature of the fluid, is a
perfect gradient (for isentropic flow $\nabla p / \rho_0 = \nabla w $).  Hence
the first integral vanishes, and the circulation is indeed time independent as
required by our theorem.

As a second example consider an axisymmetric differentially
rotating fluid exhibiting a purely poloidal magnetic field. Let the
flow also be isentropic and stationary. We choose to work in
cylindrical coordinates $\{\varrho,\phi,z\}$; the hat symbol will denote a
unit vector in the stated direction.  It then follows that
$\rho=\rho_0({\bf r})$, ${\bf B} = {\bf B}_0({\bf r})$ and ${\bf
  v}=\Omega \varrho \hat{\phi}$, where $\Omega (\varrho,z)$ is the angular
velocity of the fluid. It is well known \cite{chandra,mestel}
that for axisymmetric fields the curl of a poloidal field is a toroidal
one, and the toroidal field has only a $\hat{\phi}$ component. Therefore, the
electric current density ${\bf J} = \nabla \times {\bf B}/(4 \pi)$, is
everywhere collinear with ${\bf v}$ and time independent.  Since the
problem is stationary, $\Omega$ satisfies Ferraro's \cite{ferraro,mestel} law 
of isorotation ${{\bf B} \cdot \nabla \Omega = 0}$. In addition the
field must be torque-free \cite{mestel} i.e. no Lorentz force in the
$\hat{\phi}$ direction.  This condition is identically satisfied for  a purely
poloidal field. Combining all of the above we get the following solution of
Eq.~(\ref{dR}):
\begin{equation}
\label{ex_1}
{\bf R}=-t  {\bf J}
\end{equation}

According to Eq.~(\ref{mke1}) the conserved circulation should be
\begin{equation}
\label{ex_2}
\Gamma= \oint _{{\cal C}} \Omega
\varrho^2 d \phi  -t\oint _{{\cal C}} 
{{\bf J} \times{\bf B}_0\over
4\pi\rho_0}\cdot d{\bf r}
\end{equation}
where we have exploited the axisymmetry to rewrite the first term.
We now verify that this circulation is indeed conserved.  Because of the
differential rotation, the contour ${\cal C}$ is gradually deformed in the
azimuthal direction.  The difference $d \phi$ in the azimuthal
coordinates between two infinitesimally close fluid elements lying on
$\cal C$ can be written as
$d\phi = d\phi_0 + t\, d\Omega$ where $d\phi_0$ is the initial difference
in azimuthal coordinates while $d \Omega $ is the difference between the
elements' angular velocities. Hence we have
\begin{equation}
\label{ex_4}
\oint _{{\cal C}} \Omega
\varrho^2 d \phi  = \oint _{{\cal C}} \Omega
\varrho^2 d \phi_0 +t\oint _{{\cal C}} \Omega
\varrho^2 d \Omega.
\end{equation}
Note that the first term is time independent while the second one is linear in
time. 

The magnetic Euler equation (\ref{mag_euler}) in cylindrical coordinate reads
\begin{equation}
\label{ex_5}
-{\Omega ^ 2}\varrho \hat \varrho = -{\nabla p \over \rho_0} - \nabla U + 
{{\bf J}
 \times
  {\bf B}_0 \over 4 \pi \rho_0}.
\end{equation}
Again, by the isentropic condition we can write $\nabla p/\rho_0 =\nabla w$. 
Taking the integral round $\cal C$ of both sides of Eq.~(\ref{ex_5}) we have
\begin{equation}
\label{ex_6}
 -\oint_{{\cal C}} {\Omega ^ 2}\varrho d\varrho =\oint _{{\cal C}} {{\bf J}
\times{\bf B}_0\over
  4\pi\rho_0}\cdot d{\bf r}.
\end{equation}
Substituting from Eq.~(\ref{ex_6}) and Eq.~(\ref{ex_4}) into Eq.~(\ref{ex_2})
we get
\begin{eqnarray}
\label{ex_gamma}
\Gamma &=& \oint _{{\cal C}} \Omega \varrho^2 d \phi_0 + t\oint _{{\cal C}}
\Omega \varrho^2 d \Omega + t \oint_{{\cal C}}  {\Omega ^2}\varrho d\varrho
\nonumber
\\
&=& \oint _{{\cal C}} \Omega \varrho^2 d \phi_0 +{t \over 2} \oint _{{\cal C}}
d(\Omega^2 \varrho^2)
        = \oint _{{\cal C}} \Omega \varrho^2 d \phi_0  ,
\end{eqnarray}
and $\Gamma$ is indeed time independent.
Note that it is possible to add to ${\bf R}$ in Eq.~(\ref{ex_1}) an arbitrary
time independent solenoidal vector field ${\bf  R}_0({\bf r})$ which satisfies
${\bf  R}_0 \times {\bf v} = \nabla \chi $. However, as already stressed in the
previous subsection, this will only add to  $\Gamma$  a time independent
quantity.

It is important to note that although the example specifically relates
to an axisymmetric problem, Eq.~(\ref{ex_1}) applies to all stationary MHD 
flows which have  ${\bf J}$ collinear with ${\bf v}$. Accordingly,
$\Gamma$ will be conserved in all such flows.

\section{Relativistic Variational principle}

In this section we formulate a Lagrangian density for MHD flow
in the framework of general relativity (GR). 
Greek indices run from 0 to 3.  The coordinates are denoted
$x^\alpha=\left(x^0,x^1,x^2,x^3\right)$; $x^0$ stands for time.  A comma
denotes  the usual partial derivative; a semicolon covariant
differentiation.   Our signature is $\{-,+,+,+\}$. We continue to take $c=1$.

\subsection{Relativistic MHD Equations}
The general relativistic (GR) equations for MHD were formulated by
Lichnerowicz\cite{lichnerowicz}, Novikov  and  Thorne\cite{novikov},
Carter\cite{carter1}, Bekenstein  and  Oron\cite{bekenstein1} and others.  The
role of the mass conservation equation (\ref{continu}) is taken over by
the law of particle number conservation,
\begin{equation}
\label{gr4}
N^\alpha{}_{;\alpha }=\left( nu^{\alpha }\right) _{;\alpha }=0,
\end{equation}
where $N^\alpha$ is the particle number 4--current density, $n$ the particle
proper number density and $u^\alpha$ the fluid 4--velocity field normalized by
$u^\alpha u_\alpha=-1$.  If $s$ represents the entropy per particle (not per
unit mass as in Sec.~II), then the assumption of ideal adiabatic flow,
Eq.~(\ref{adiabatic0}),  can be put in the form
\begin{equation}
\label{gr_entropy}
\left( sN^{\alpha }\right) _{;\alpha }=0\qquad {\rm or} \qquad u^\alpha
s_{,\alpha}=0.
\end{equation}

Because the flow is assumed adiabatic, the energy momentum tensor for the
magnetized fluid is that of an ideal fluid augmented by the electromagnetic
energy--momentum tensor:
\begin{equation}
\label{gr5}
T^{\alpha \beta }=pg^{\alpha \beta }+\left( p+\rho \right) u^{\alpha }
u^{\beta }+  (F^{\alpha\gamma} F^\beta{}_\gamma - {\scriptstyle 1\over
\scriptstyle 4} F^{\gamma\delta} F_{\gamma\delta}\, g^{\alpha\beta})/(4\pi).
\end{equation}
Here $\rho$ represents the fluid's energy proper density (including rest
masses) and $p$ the scalar pressure (again assumed isotropic), while
$F^{\alpha\beta}$ denotes the electromagnetic field tensor.  As usual the
covariant divergence
$T^{\alpha \beta }{}_{;\beta}$ must vanish (energy--momentum conservation). In
consequence
$T^{\alpha \beta}{}_{;\beta} +u^\alpha u_\gamma T^{\gamma \beta
}{}_{;\beta} = 0 $ which can be recast as
\begin{equation}
\label{em_conservation}
(\rho+p)u^\beta u^\alpha{}_{;\beta} = - (g^{\alpha\beta}+u^\alpha
u^\beta)p_{,\beta} + F^{\alpha\beta} F_{\beta}{}^\gamma{}_{;\gamma}/(4\pi).
\end{equation}
The term $a^\alpha\equiv u^\beta u^\alpha{}_{;\beta}$ stands for the fluid's
acceleration 4--vector.  The effects of gravitation are automatically
included by the appeal to curved metric and covariant derivatives. This
equation parallels  Eq.~(\ref{mag_euler});as usual in GR the
pressure contributes alongside the energy density to the inertia. The
electromagnetic field tensor is subject to Maxwell's equations
\begin{eqnarray}
\label{Maxwell1}
F^{\alpha\beta}{}_{;\beta} &=& 4\pi j^\alpha
\\
 F_{\alpha\beta,\gamma} +
F_{\beta\gamma,\alpha}+F_{\gamma\alpha,\beta}  &=& 0.
\label{Maxwell2}
\end{eqnarray}
where $j^\alpha$ denotes the electric 4--current density.
Putting all this together we have the GR MHD Euler equation
\begin{equation}
\label{grmag_euler}
(\rho+p)a^\alpha = - h^{\alpha\beta} p_{,\beta} +
F^{\alpha\beta} j_\beta,
\end{equation}
where we have introduced the projection tensor
\begin{equation}
\label{h}
 h^{\alpha\beta}\equiv g^{\alpha\beta} + u^\alpha u^\beta.
\end{equation}

The above equations do not completely specify MHD flow (as opposed to flow
of a generic magnetofluid).  For any flow carrying an electromagnetic field,
the (antisymmetric) Faraday tensor $F_{\alpha \beta }$ may be split into
electric and magnetic vectors with respect to the flow:
\begin{eqnarray}
\label{gr1}
E_{\alpha }&=&F_{\alpha \beta }u^{\beta }
\\
\label{gr2}
B_{\alpha }&=&{}^*F_{\beta\alpha}u^\beta\equiv {\scriptstyle
1\over\scriptstyle 2}\epsilon_{\beta\alpha
\gamma \delta }\,F^{\gamma \delta } u^{\beta }.
\end{eqnarray}
Here $ \epsilon _{\alpha \beta \gamma \delta } $ is the
Levi-Civita totally antisymmetric tensor ($\epsilon_{0123}=(-g)^{1/2}$ with
$g$ denoting the determinant of the metric $g_{\alpha\beta}$) and
${}^*F_{\alpha\beta}$ is the dual of $F_{\alpha\beta}$.  In the frame moving
with the fluid, these 4--vectors have only spatial parts which correspond to
the usual ${\bf E}$ and ${\bf B}$, respectively.  The inversion of
Eqs.~(\ref{gr1}-{\ref{gr2}) is
\begin{equation}
F_{\alpha\beta}=u_\alpha E_\beta - u_\beta E_\alpha
+\epsilon_{\alpha\beta\gamma\delta}u^\gamma B^\delta
\label{inversion}
\end{equation}  
For  an infinitely conducting (perfect
MHD) fluid, the electric field in the fluid's frame must vanish, i.e., 
\begin{equation}
\label{gr3}
E_{\alpha }=F_{\alpha \beta } u^\beta = 0.
\label{freeze}
\end{equation}
This corresponds to the usual MHD condition  ${\bf E} + {\bf v} \times {\bf
B}=0$.

\subsection{Relativistic Lagrangian Density }
\label{rel_lagrangian}

Inspired by Schutz's\cite{schutz} Lagrangian density for
pure fluids in GR, we now propose a Lagrangian density for GR MHD
flow which reproduces Eqs.~(\ref{gr4}-\ref{gr_entropy}),
(\ref{Maxwell1}-\ref{grmag_euler}) and (\ref{gr3}).  Like Schutz we include
Lin's term, which proves essential to our subsequent proof of the existence
of a circulation theorem.  The proposed Lagrangian density is
\begin{equation}
\label{rl_1}
{\cal L}=-\rho(n,s) -F_{\alpha \beta }F^{\alpha \beta }/(16\pi)+
\phi N^{\alpha }{}_{;\alpha }+\eta \left( sN^{\alpha }\right) _{;\alpha
}+\lambda \left( \gamma N^{\alpha }\right) _{;\alpha }+\tau^{\alpha
}F_{\alpha\beta }N^{\beta }.
\label{Lagrangian}
\end{equation}
Now in GR the scalar density ${\cal L}\,(-g)^{1/2}$ replaces ${\cal L}$
in the action (\ref{action}), and is what enters in the Euler--Lagrange
equations (\ref{EL}).  The covariant derivatives cause no problem; for
example $(-g)^{1/2}\phi N^\alpha{}_{;\alpha}
=\phi[(-g)^{1/2}N^\alpha]_{,\alpha}$, whose variation with respect to
$N^\alpha$ is easily integrated by parts.

As in the nonrelativistic case, $\phi $ is the Lagrange multiplier
associated with the conservation of particle number constraint,
Eq.~(\ref{gr4}),   $
\eta  $ is that multiplier associated with the adiabatic flow constraint,
Eq.~(\ref{gr_entropy}), and $\lambda  $ is  that associated with the
conservation along the flow of Lin's quantity $\gamma$.  We view $\gamma$,
$N^\alpha$ and $s$ as the independent fluid variables, while $n$ and
$u^\alpha$ are determined  by the obvious relations
\begin{equation}
\label{grem0}
-N^{\alpha }N_{\alpha }=n^2; \qquad u^\alpha = n^{-1}\,N^\alpha.
\end{equation}
Strictly speaking, one should include in ${\cal L}$ a new Lagrange
multiplier times the constrained expression $N^{\alpha }N_{\alpha }+n^2$.
Rather than clutter up ${\cal L}$ further, we enforce this constraint
below by hand. 

As usual, we view the vector potential $A_\alpha$, rather than the
electromagnetic field tensor $F_{\alpha\beta} = A_{\beta;\alpha} -
A_{\alpha;\beta} = A_{\beta,\alpha} -A_{\alpha,\beta}$, as the independent
electromagnetic variable.   In consequence, the Maxwell Eqs.~(\ref{Maxwell2})
are satisfied as identities. The last term in ${\cal L}$ enforces the
``vanishing of electric field''  constraint, Eq.~(\ref{gr3}); $ \tau^{\alpha }
$ is a Lagrange multiplier 4--vector field.  Because here we enforce the
``vanishing of electric field'' rather than the equivalent flux freezing
condition (\ref{ff}), the $\tau^\alpha$ is more like ${\bf R}$ of Sec.~II.B.
than like ${\bf K}$.  Not all of $\tau^\alpha$ is physically meaningful.  For
suppose we add an arbitrary function $f(x^\beta)$ multiplied by $N^\alpha$ to
$\tau^\alpha$.  This increments the Lagrangian density  by $f
N^\alpha F_{\alpha\beta} N^\beta$ which vanishes identically by the
antisymmetry of $F_{\alpha\beta}$.   So $\tau^\alpha$ and $\tau^\alpha+f
N^\alpha$ are physically equivalent.  We shall exploit this to substract from
$\tau^\alpha$ its component along $u^\alpha$.  So henceforth we may take it
that $\tau^\alpha u_\alpha=0$.  

Much freedom is still left in $\tau^\alpha$. Suppose we add
to it a term proportional to $n^{-1}\,B^\alpha$.  By
Eqs.~(\ref{gr1}-\ref{inversion}), this adds to the Lagrangian density the
term $ E_\alpha B^\alpha$.  Of course we cannot take this to
vanish at the Lagrangian's level because we have not yet obtained the
freezing-in condition (\ref{freeze}) from it.  However, it is known that
$B^\alpha E_\alpha={\scriptstyle 1\over\scriptstyle
4}\epsilon^{\alpha\beta\gamma\delta} F_{\alpha\beta} F_{\gamma\delta}$.  By
introducing the potential $A_\alpha$ we can write this as $ {\scriptstyle
1\over\scriptstyle 2}\left[\epsilon^{\alpha\beta\gamma\delta} F_{\alpha\beta}
A_{\gamma}\right]_{;\delta} - {\scriptstyle 1\over\scriptstyle
2}\epsilon^{\alpha\beta\gamma\delta}  F_{\alpha\beta;\delta} A_\gamma$,
where we have used the fact that
$\epsilon^{\alpha\beta\gamma\delta}$ has vanishing covariant derivatives. 
Obviously the last term vanishes by virtue of Maxwell's equations
(\ref{Maxwell2}) which are identities in the present approach.  When
multiplied by $(-g)^{1/2}$, the first term becomes a perfect
derivative.  Such term, when added to the integral forming the Lagrangian,
is known not to affect its physical content.  Thus $\tau^\alpha$ and
$\tau^\alpha+{\rm const.}\hskip-3pt\times n^{-1}\,B^\alpha$ are physically
equivalent, and this transformation respects the condition $\tau_\alpha
u^\alpha=0$ because $B_\alpha u^\alpha=0$ [see (\ref{gr2})].    However,
there is not enough freedom in the constant to allow us to eliminate the
component of $\tau^\alpha$ along $B^\alpha$.  But in Sec.\ref{last_sec} we
shall exploit what we have just found.  

\subsection{Equations of Motion}

We can now derive the equations of motion. Variation of $\phi$
recovers the conservation of particles $N^\alpha{}_{;\alpha}$.  Variation of
$\lambda$ with subsequent use of the previous result yields
\begin{equation}
\gamma_{,\alpha }u^{\alpha }=0.
\label{gr_lambda}
\end{equation}
If we vary $ \gamma  $ we get
\begin{equation}
\lambda_{,\alpha }u^{\alpha }=0.
\label{gr_gamma}
\end{equation}
These two results are precise analogs of Eqs.~(\ref{Lin}) and (\ref{le4});
they inform us that $\gamma$ and $\lambda$ are both locally conserved with
the flow.  In view of the thermodynamic relation
$n^{-1}(\partial \rho/\partial s)_n = T$, with $T$ the locally
measured fluid temperature, variation  of
$s$ gives
\begin{equation}
u^{\alpha }\eta _{,\alpha }=-T;
\label{gr_s}
\end{equation}
this is the analogue of Eq.~(\ref{le6}).

We now vary $N^\alpha$ using the obvious consequence of Eq.~(\ref{grem0}),
\begin{equation}
\label{grem2}
\delta n=-u_{\alpha }\delta N^{\alpha },
\end{equation}
together with the thermodynamic relation\cite{novikov} involving the
specific enthalpy $\mu$,
\begin{equation}
\label{gr_mu1}
\mu \equiv (\partial\rho/\partial n)_s =n^{-1}\,(\rho +p);
\end{equation}
we get the GR analog of Eq.~(\ref{le3}),
\begin{equation}
\mu u_{\alpha }=\phi _{,\alpha }+s\eta _{,\alpha }+\gamma
\lambda _{,\alpha }+
\tau^{\beta }F_{\alpha\beta }.
\label{gr_varu}
\end{equation}
The importance of Lin's $\gamma$ is again clear here; in the pure
isentropic fluid case ($F^{\alpha\beta}=0$ and $s=$ const.), the
Khalatnikov vorticity tensor given by
\begin{equation}
\label{gr_vor2}
\omega _{\alpha \beta }=\left( \mu u_{\beta }\right) _{,\alpha }-
\left( \mu u_{\alpha }\right) _{,\beta }
=\left( \gamma \lambda _{,\beta }\right) _{,\alpha }-
\left( \gamma \lambda _{,\alpha }\right) _{,\beta }
\end{equation}
would vanish in the absence of $\gamma$, thus constraining us to discuss
only irrotational flow.

By contracting Eq.~(\ref{gr_varu})  with $u^\alpha$ and using $u_\alpha
u^\alpha=-1$ as well as Eqs.~(\ref{gr3}) and
(\ref{gr_gamma}-{\ref{gr_s}), we get the following GR version of
Eq.~(\ref{le7}):
\begin{equation}
\phi_{,\alpha}u^\alpha = - \mu + Ts.
\label{gr_phi}
\end{equation}
Thus the proper time rate of change of $\phi$ along the flow is just minus the
specific Gibbs energy or chemical potential.  The apparent difference
between the result here and Eq.~(\ref{le7}) stems from the fact that proper
time rate (here) and coordinate time rate (there) differ by gravitational
redshift and time dilation effects.  These effects are not noticeable when
one compares Eqs.~(\ref{gr_s}) with (\ref{le6}) because the first refers to
locally measured temperature and the second to global temperature; these two
temperatures differ by the same factors as do proper and coordinate time.

Turn now to the variation of $A_\alpha$.  Because of the antisymmetry of
$F_{\alpha\beta}$, the last term of the Lagrangian, Eq.~(\ref{rl_1}), can
be written as $(\tau^\beta N^\alpha-\tau^\alpha N^\beta)A_{\alpha,\beta}$.
The variation of $A_\alpha$ in the corresponding term in the action
produces, after integration by parts, the term $\big[(-g)^{1/2}(\tau^\alpha
N^\beta-\tau^\beta N^\alpha)\big]_{,\beta}\, \delta A_\alpha$.  Because
for any antisymmetric tensor $t^{\alpha\beta}$,
$(-g)^{1/2}t^{\alpha\beta}{}_{;\beta} = [(-g)^{1/2}
t^{\alpha\beta}]_{,\beta}$, this finally leads to the equation
\begin{equation}
\label{gr_mag2}
F^{\alpha \beta }{}_{;\beta } =4\pi\left( \tau^{\alpha }N^{\beta }
- \tau^{\beta }N^{\alpha }\right) _{;\beta }.
\label{tau}
\end{equation}
Comparison with Eq.~(\ref{Maxwell1}) shows that this result gives us a
representation of the electric current density $j^\alpha$ as the
divergence of the bivector $ \tau^{\alpha }N^{\beta }- \tau^{\beta }
N^{\alpha}$.  Such representation makes the conservation of charge
($j^\alpha{}_{;\alpha}=0) $ an identity  because the divergence of the
divergence of any antisymmetric tensor vanishes. This equation is the GR
analogue of Eq~(\ref{dR}). Formally Eq.~(\ref{tau}) determines the Lagrange 
multiplier 4--vector $\tau^\alpha$, modulo the freedom inherent in it.

\subsection{MHD Euler Equation in General Relativity}

Our central task now is to show that the equations in Sec.~III.C
lead uniquely to the GR MHD Euler equation (\ref{grmag_euler}). We begin by
writing the Khalatnikov vorticity $\omega_{\beta\alpha}$ in two forms,
\begin{equation}
\omega_{\beta\alpha} = \mu_{,\beta} u_\alpha - \mu_{,\alpha} u_\beta
 +\mu u_{\alpha;\beta} - \mu u_{\beta;\alpha},
\end{equation}
as well as by means of Eq.~(\ref{gr_varu})
\begin{eqnarray}
\label{gr_mgel2}
\omega_{\beta\alpha} &=& s_{,\beta }\eta _{,\alpha }-s_{,\alpha }\eta
_{,\beta}+\gamma _{,\beta }\lambda _{,\alpha }-\gamma _{,\alpha }\lambda
_{,\beta }
 \nonumber \\
&+&\tau^{\delta }{}_{;\beta }F_{\alpha \delta }-\tau^{\delta }{}_{;\alpha
}F_{\beta\delta }+\tau^{\delta }F_{\alpha \delta ;\beta }-\tau^{\delta
}F_{\beta
\delta;\alpha }.
\end{eqnarray}
Contracting the left hand side of the first with $ N^{\alpha } $, recalling
Eq.~(\ref{grem0}) and that by normalization $ u^{\alpha }u_{\alpha ;\beta
}=0 $ whereas $ u^{\beta }u_{\alpha ;\beta }=a_\alpha $, the fluid's
4--acceleration, we get
\begin{equation}
\label{gr_mgel3}
\omega_{\beta\alpha} N^\alpha =-n\mu _{,\beta }-n\mu _{,\alpha }u^{\alpha
}u_{\beta }-n\mu a_\beta = -n h_\beta{}^\alpha\mu_{,\alpha} - n \mu a_\beta.
\end{equation}

Now contracting Eq.~(\ref{gr_mgel2}) with  $ N^{\alpha } $ and using
Eqs.~(\ref{gr_lambda}-\ref{gr_s}) and (\ref{gr3}) to drop a number of terms
we get
\begin{equation}
\label{gr_mgel6}
\omega_{\beta\alpha} N^\alpha =-nTs_{,\beta }-\tau^{\delta }{}_{;\alpha
}F_{\beta \delta }N^{\alpha }+\tau^{\delta }F_{\alpha \delta ;\beta
}N^{\alpha}-\tau^{\delta }F_{\beta \delta ;\alpha }N^{\alpha }.
\end{equation}
By virtue of Eq.~(\ref{gr_entropy}), $-nTs_{,\beta}$ is the same as
$-n T h_\beta{}^\alpha s_{,\alpha}$.  It is convenient  to use
the thermodynamic identity $d\mu =n^{-1}\,dp+Tds$, which follows from
Eq.~(\ref{gr_mu1}) and the first law $d(\rho/n)= Tds -pd(1/n)$, to replace
in Eq.~(\ref{gr_mgel6})  $-nTs_{,\beta}$ by
$h_\beta{}^\alpha (-n\mu_{,\alpha}+p_{,\alpha})$.  Equating our two
expressions for $\omega_{\beta\alpha} N^\alpha$ gives, after a cancellation,
\begin{equation}
\label{gr_mgel9}
-\left( n\mu a_{\beta }+h_{\beta }{}^{\alpha }p_{,\alpha }\right) =
-\tau^{\delta }{}_{;\alpha }F_{\beta \delta }N^{\alpha }+\tau^{\delta
}F_{\alpha\delta ;\beta }N^{\alpha }-\tau^{\delta }F_{\beta \delta ;\alpha
}N^{\alpha }.
\end{equation}

The last two terms in this equation can be combined into a single one by
virtue of Eq.~(\ref{Maxwell2}), which, as well known, can be written with
covariant as well as ordinary derivatives.  Further, by Eq.~(\ref{gr_mu1})
we may replace
$n\mu$ by $\rho+p$.  In this manner we get
\begin{equation}
\left( \rho +p\right) a_{\beta }=-h_{\beta }{}^{\alpha }p_{,\alpha }+
F_{\beta \alpha ;\delta }\tau^{\delta }N^{\alpha }+F_{\beta \delta
}\tau^{\delta }{}_{;\alpha}N^{\alpha }.
\end{equation}
The term $ \tau^{\delta }{}_{;\alpha }N^{\alpha } $ here can be replaced by
two others with help of Eq.~(\ref{gr_mag2}) if we take into
account that $N^\beta{}_{;\beta}=0$:
\begin{equation}
\label{gr_mgel12}
\left( \rho +p\right) a_{\beta }=-h_{\beta }{}^{\alpha }p_{,\alpha }+
F_{\beta \delta }F^{\delta \alpha }{}_{;\alpha }/(4\pi)+F_{\beta \alpha
;\delta }\tau^{\delta }N^{\alpha }+F_{\beta \delta }\left( \tau^{\alpha
}N^{\delta}\right) _{;\alpha }.
\end{equation}
We note that the last two terms on the right hand side combine into $\left(
F_{\beta \alpha }N^{\alpha }\tau^{\delta }\right) _{;\delta }$ which
vanishes by Eq.~(\ref{gr3}). Now substituting from the
Maxwell equations (\ref{Maxwell1}) we arrive at the final equation
\begin{equation}
\label{gr_mgel15}
\left( \rho +p\right) a_{\beta }=-h_{\beta }{}^{\alpha }p_{,\alpha }+
F_{\beta \delta }j^{\delta },
\end{equation}
which is the correct GR MHD Euler equation (\ref{grmag_euler}).  We have not
used any information about $\tau^\alpha$ beyond Eq.~(\ref{tau}); hence
Euler's equation is valid for all choices of $\tau^\alpha$.  Since we
are able to obtain all equations of motion for GR MHD from our Lagrangian
density, we may regard it as correct, and go on to look at some consequences.

\subsection{New Circulation Conservation Law}
\label{last_sec}

Eqs.~(\ref{gr_varu}) and (\ref{gr_lambda}-\ref{gr_gamma}) allow us to
generalize the conserved circulation of Sec.~II.D to relativistic perfect
MHD.   Let
$ \Gamma $ be the line integral
\begin{equation}
\label{gr_kel1}
\Gamma =\oint_{{\cal C}}z_{\alpha }dx^{\alpha },
\end{equation}
where $ {\cal C} $ is, again, a closed curve drifting with the fluid, and
\begin{equation}
\label{gr_kel2}
z_\alpha \equiv {\mu}u_{\alpha }-\tau^{\beta }F_{\alpha \beta }.
\end{equation}
According to Eq.~(\ref{gr_varu}), $ z_{\alpha }=\phi _{,\alpha }+s\eta
_{,\alpha }+\gamma \lambda _{,\alpha }$.   Since $ \phi _{,\alpha } $ is a
gradient, its contribution to  $\Gamma $ vanishes. Likewise, for isentropic
flow ($ s=$ const.) the term involving $ s\eta _{,\alpha } $ makes no
contribution to $\Gamma$. Thus
\begin{equation}
\label{gr_kel3}
\Gamma =\oint_{{\cal C}}\gamma \lambda_{,\alpha} dx^\alpha = \oint_{{\cal
C}}\gamma\, d\lambda.
\end{equation}
By Eqs.~(\ref{gr_lambda}-\ref{gr_gamma}) both $\gamma$  and $ \lambda$ are
conserved with the flow.  Thus $ \Gamma  $ is conserved along the flow. 
Note that by virtue of $\gamma$'s presence, $\Gamma$ need not vanish. 

In the absence of electromagnetic fields and in the nonrelativistic limit
$(\mu\rightarrow m$ where $m$ is a fluid particle's rest mass), $\Gamma $
for a curve ${\cal C}$ taken at constant time reduces to Kelvin's
circulation.  On this ground our result can be considered a generalization
of Kelvin's circulation theorem to general relativistic
MHD. We have gone here beyond Oron's original result\cite{bekenstein1} in
that no symmetry is necessary for the circulation to be conserved.

To manifestly exhibit the conserved circulation, one has to know
$\tau^\alpha$ explicitly. The first step is to understand the
freedom left in $\tau^\alpha$ beyond that discussed in
Sec.~\ref{rel_lagrangian}.  The second is to determine
$\tau^\alpha$ in a specific flow exploiting for this the symmetries and
other information.  Below we address the first step; the second is left
mainly to future publications.  

Given a specific MHD flow as background, let us at define a generic test field
$f_{\alpha\beta}=-f_{\beta\alpha}$ which satisfies Maxwell's homogeneous
equations (\ref{Maxwell2}) as well as the freezing-in condition
(\ref{freeze}), e.g.  $e_\alpha \equiv f_{\alpha\beta}u^\beta=0$.   We think
of $f_{\alpha\beta}$ as very weak, so that it does not disturb the MHD flow
or the spacetime geometry; it is a passive tensor.  Because
$f_{\alpha\beta}u^\beta=0$, $f_{\alpha\beta}$ has only three independent
components.  Therefore, its full content is reflected in the ``magnetic
4-vector'' $b_{\alpha}\equiv{\scriptstyle 1\over\scriptstyle
2}\epsilon_{\beta\alpha \gamma \delta }\, f^{\gamma \delta }u^{\beta }$,
which is obviously orthogonal to $u_\alpha$.  The transformation
$\tau^\alpha\rightarrow \tau^\alpha +k n^{-1}\,b^\alpha$ ($k$ a real
constant) is not a symmetry of the Lagrangian.  However, it does not disturb
the inhomogeneous Maxwell equations (\ref{Maxwell1},\ref{tau}).  This is
because the change in $\tau^\alpha$ merely adds to the electric current the
term $\left( b^{\alpha }u^{\beta } - b^{\beta }u^{\alpha }\right)_{;\beta
}=(-g)^{-1/2}\left[ (-g)^{1/2} (b^{\alpha }u^{\beta } - b^{\beta }u^{\alpha
})\right]_{,\beta }$.  Because of the condition $e^\alpha = 0$, we may easily
invert the analog of (\ref{inversion}) to get $b^{\alpha }u^{\beta } -
b^{\beta }u^\alpha = {\scriptstyle 1\over \scriptstyle
2}\epsilon^{\alpha\beta\gamma\delta} f_{\gamma\delta}$.  But since
$(-g)^{1/2}\epsilon^{\alpha\beta\gamma\delta}$ is just the constant
antisymmetric symbol, our assumed equations $f_{\alpha\beta,\gamma}
+f_{\gamma\alpha,\beta}+f_{\beta\gamma,\alpha}=0$ imply that $\left(
b^{\alpha }u^{\beta } - b^{\beta }u^{\alpha }\right)_{;\beta}=0$ so that the
Maxwell equations (\ref{tau}) are invariant under $\tau^\alpha\rightarrow
\tau^\alpha +k n^{-1}\,b^\alpha$.  So is the Euler
equation since its derivation used only the information about
$\tau^\alpha$ inherent in (\ref{tau}). 

The expression for $u_\alpha$, (\ref{gr_varu}), does seem to change under
$\tau^\alpha\rightarrow \tau^\alpha +k n^{-1}\,b^\alpha$, and we also note
that $\Gamma\rightarrow \Gamma+k \oint n^{-1}\,b^\beta F_{\alpha\beta}\,
dx^\alpha$.  Now since the ``magnetic 4-vector'' $b^\alpha$ is frozen in,
like any such {\it infinitesimal\/} field, it evolves in such a way that
$n^{-1}\,b^\alpha$ gives for all time that part of the spacetime separation
of two neighboring fluid elements which is orthogonal to
$u^\alpha$\cite{bekenstein1}, {\it c.f.\/} discussion after
Eq.~(\ref{DGamma}).  Thus
$n^{-1}\,b^\alpha$ can be employed to define a thin closed strip dragged with
the fluid such that one of its edges coincides with the curve ${\cal C}$. 
Therefore, the increment $\oint n^{-1}\,b^\beta F_{\alpha\beta}\, dx^\alpha$
is just the conserved magnetic flux through this strip.  Evidently the
transformation $\tau^\alpha\rightarrow \tau^\alpha +k n^{-1}\,b^\alpha$
has not changed the nature of the conservation law for $\Gamma$, but only
added a trivially conserved quantity to it..  

Now the MHD flow $\{ B^\alpha, u^\alpha, n, \rho, \mu\}$ is evidently
unchanged because neither the MHD Euler equation (\ref{em_conservation}) nor
Maxwell's equations were changed, so we must conclude that in the expression
for $u^\alpha$, Eq.~(\ref{gr_varu}), the change of the $\tau^\beta
F_{\alpha\beta}$ term must be compensated by suitable changes in the pair of
Lagrange multipliers $\phi+s\eta$ and $\lambda$ (since we are assuming $s={\rm
const.}$).  They are capable of this because $b^\alpha$ has only two
independent components.  For the condition $b^\alpha u_\alpha=0$ eliminates
one of the four. In addition  $b^\alpha$ comes from $f_{\alpha\beta}$ which
satisfies Eqs.~(\ref{Maxwell2}); in particular, $f_{12,3}+f_{31,2}+f_{23,1}=0$
in the chosen coordinates.  But since no time derivatives appear in it, this
last equation serves as an initial constraint on $b^\alpha$ just as
the Gauss equation $\nabla\cdot{\bf B}=0$ does for the true magnetic field. 
Accordingly, one further relation exists between components of $b^\alpha$ so
that the generic $b^\alpha$ contains only two free functions.  Thus the
change in $\tau^\beta F_{\alpha\beta}$ can be taken up by changes in the two
functions $\phi+s\eta$ and $\lambda$ so that $\mu u_\alpha$ is unchanged. 
 
Note that it is not possible to ``get rid'' of $\tau^\alpha$ by means of
the transformation $\tau^\alpha\rightarrow \tau^\alpha +k n^{-1}\,b^\alpha$ 
because, as we shall make clear presently, $\tau^\alpha$ and $b^\alpha$ obey
different equations of motion.  Thus there must be a residual part of
$\tau^\alpha$ which is not changed by the transformations.  It is this part
which is responsible for the conserved circulation, so that the conservation
of $\Gamma$ cannot be reduced to magnetic flux conservation.

The following algorithm can be used to find $\tau^\alpha$.  Maxwell's
inhomogeneous equations (\ref{tau}) which say that the divergence of a
certain tensor vanishes can always be ``solved'' by the prescription
\begin{equation}
F^{\alpha\beta}-4\pi(\tau^\alpha N^\beta-\tau^\beta N^\alpha)={\scriptstyle
1\over\scriptstyle 2}\epsilon^{\alpha\beta\gamma\delta}{\cal F}_{\gamma\delta}
\label{prescription}
\end {equation}
where the new field ${\cal F}_{\gamma\delta}$ just has to satisfy Maxwell's
homogeneous equations (\ref{Maxwell2}), i.e. ${\cal F}_{\gamma\delta}\equiv
{\cal A}_{\delta,\gamma}-{\cal A}_{\gamma,\delta}$. Taking the dual of
Eq.~(\ref{prescription}) with help of the identity
$\epsilon_{\gamma\delta\alpha\beta}\epsilon^{\alpha\beta\mu\nu}=
-2(\delta_\gamma{}^\mu\delta_\delta{}^\nu
-\delta_\gamma{}^\nu\delta_\delta{}^\mu)$ gives
\begin{equation}
{}^*F_{\gamma\delta}-4\pi\epsilon_{\gamma\delta\alpha\beta} \tau^\alpha
N^\beta=-{\cal F}_{\gamma\delta}
\label{dual}
\end {equation}
Contracting this equation with $u^\gamma$ gives the further requirement on
${\cal F}_{\alpha\beta}$:
\begin{equation}
{\cal F}_{\delta\gamma} u^\gamma=B_\delta,
\label{requirement}
\end {equation}
where we have used Eq.~(\ref{gr2}).  The ${\cal F}_{\delta\gamma}$ can
always be solved for: because of gauge freedom there are three independent
components in ${\cal A}_\alpha$, and this is enough to find a solution for
an arbitrary field $B_\delta$ obeying $B_\alpha u^\alpha=0$ (thus three
components at most). If fact, $B_\delta$ does not determine
${\cal F}_{\delta\gamma}$ uniquely: if one adds to this last one of the
frozen-in
$f_{\gamma\delta}$ we discussed earlier in this section (which also satisfy
the homogeneous Maxwell equations), Eq.~(\ref{requirement}) is still
satisfied because $f_{\delta\gamma}u^\gamma=0$.  

We get $\tau^\alpha$ by contracting Eq.~(\ref{prescription}) by $u_\beta$ and
remembering that $F^{\alpha\beta}u_\beta=0$ and $\tau^\beta u_\beta=0$.  Thus
\begin{equation}
\tau^\alpha=(8\pi n)^{-1}\epsilon^{\alpha\beta\gamma\delta}{\cal
F}_{\gamma\delta}\,u_\beta
\end{equation}
It is interesting that $B_\delta$ plays the role of electric part of ${\cal
F}_{\delta\gamma}$ while $\tau^\alpha$ enters like the magnetic part of this
tensor, {\it c.f.\/} Eq.~(\ref{gr2}) (but because ${\cal F}_{\delta\gamma}
u^\gamma\neq 0$, $\tau^\alpha$ evolves differently from a magnetic type field
like $B^\alpha$ or the $b^\alpha$).   It should also be clear now that the
freedom in redefining ${\cal F}_{\gamma\delta}\rightarrow {\cal
F}_{\gamma\delta}+f_{\gamma\delta}$ is equivalent to the changes
$\tau^\alpha\rightarrow \tau^\alpha +k n^{-1}\,b^\alpha$ we considered
earlier in this section.  This freedom can be exploited together with the
symmetries to simplify the problem of solving explicitly for
$\tau^\alpha$ in any specific MHD flow.


\end{document}